\documentclass[conference]{IEEEtran}
\newcommand{\CLASSINPUTtoptextmargin}{2 cm}
\newcommand{\CLASSINPUTbottomtextmargin}{2.6 cm}
\IEEEoverridecommandlockouts
\usepackage{cite}

\usepackage{amsmath,amssymb,amsfonts}
\usepackage[linesnumbered,ruled,vlined]{algorithm2e}
\usepackage[dvips]{graphicx}
\usepackage{textcomp}
\usepackage{xcolor}
\usepackage[T1]{fontenc}
\usepackage{colortbl}
\usepackage{subfigure}
\usepackage{verbatim,float,caption}
\usepackage{flushend}
\flushend
\begin{document}
\title{\LARGE{Radio Resource Management and Path Planning in Intelligent Transportation Systems via Reinforcement Learning for\\ Environmental Sustainability}}
 	\author{
	\IEEEauthorblockN{S. Norouzi\IEEEauthorrefmark{1}, N. Azarasa\IEEEauthorrefmark{2}, M.R. Abedi\IEEEauthorrefmark{1}, N. Mokari\IEEEauthorrefmark{1}, S.E. Seyedabrishami\IEEEauthorrefmark{2}, H. Saeedi\IEEEauthorrefmark{3}\IEEEauthorrefmark{1}, E.A. Jorswieck\IEEEauthorrefmark{4}}
	\IEEEauthorblockA{\IEEEauthorrefmark{1}Department of Electrical and Computer Engineering, Tarbiat Modares University, Tehran, Iran} 
	\IEEEauthorblockA{\IEEEauthorrefmark{2}Department of Civil and Environmental Engineering, Tarbiat Modares University, Tehran, Iran} 
	\IEEEauthorblockA{\IEEEauthorrefmark{3}College of Engineering, Doha University for Science and Technology, Doha, Qatar}
	\IEEEauthorblockA{\IEEEauthorrefmark{4}Institute of Communications Technology, TU Braunschweig, Lower Saxony, Germany}
	\IEEEauthorblockA{Emails:\{nader.mokari, seyedabrishami\}@modares.ac.ir, hamid.saeedi@udst.edu.qa, jorswieck@ifn.ing.tu-bs.de.
	}
}
\IEEEoverridecommandlockouts
\IEEEpubid{\makebox[\columnwidth]
	{     979-8-3503-4760-9/23/\$31.00~\copyright{}2023 IEEE
		\hfill} \hspace{\columnsep}\makebox[\columnwidth]{ }}
\maketitle
\maketitle
\begin{abstract}
Efficient and dynamic path planning has become an important topic for urban areas with larger density of connected vehicles (CV) which results in reduction of travel time and directly contributes to environmental sustainability through reducing energy consumption.
CVs exploit the cellular wireless vehicle-to-everything (C-V2X) communication technology to disseminate the vehicle-to-infrastructure (V2I) messages to the Base-station (BS) to improve situation awareness on urban roads. In this paper, we investigate radio resource management (RRM) in such a framework to minimize the age of information (AoI) so as to enhance path planning results. We use the fact that 
V2I messages with lower AoI value result in less error in estimating the road capacity and more accurate path planning. Through simulations, we compare
road travel times and volume over capacity (V/C) against different levels of AoI and demonstrate the promising performance of the proposed framework.
\end{abstract}
\vspace{-0.3cm}
\section{Introduction}
Intelligent transportation systems suggest that connected vehicles (CVs) will gradually take over the roads.
Meanwhile, path planning remains an  important process to increase the driving efficiency. Path planning refers to the selection of a path from a starting point to the terminal point within a particular network under some standard criteria such as path minimization \cite{wang2019v2v}. This is specially the case in traditional path planning algorithms where the traffic volume is assumed static \cite{sunder2018}. It means that traffic jams and accidents are not taken into account.

A great deal of progress on path planning has been made in the literature
\cite{chu2015}, mainly focusing on finding the shortest  path/time \cite{enayattabar2019dijkstra, broumi2019shortest, di2022novel}.
To improve fuel consumption efficiency, a path planning scheme using deep reinforcement learning (DRL) is proposed in \cite{9102259} in which task deadline and fuel consumption of each vehicle in the platoon are considered.
 A novel three-architecture approach for vehicle path planning (SEARCH) proposed in \cite{Oubbati} which exploits Unmanned Aerial Vehicles (UAVs), Vehicular Ad hoc Networks (VANETs), 5G based cellular systems, and Software-Defined Networking (SDN) to address two main issues for navigation management difficulties: lack of local knowledge regarding navigation solutions as well as vehicles that are inflexible in dealing with unforeseeable situations. Based on their findings, the proposed architecture works effectively in terms of reducing the driving time to any destination. Authors in \cite{hui2021time}, by considering the personalized requirements of vehicle users, propose a digital-twin (DT) enabled path planning scheme to facilitate traffic management using a Q-learning algorithm. 

A key subject in intelligent transportation systems is
the vehicle communication. According to the existing literature, in \cite{Abdel2020}, we examine Mode 4 of the 3GPP cellular V2X architecture \cite{36.8851}. A distributed algorithm is used to schedule resources and manage interference between CVs. 
Since path planning information is time-sensitive, it is particularly critical to ensure the freshness of the information received by base station (BS) at intersections. 
Traditional performance metrics, such as latency or throughput, are not sufficient to measure the timeliness of updates. As such, a new criterion, called the age of information (AoI), is introduced in \cite{Molina2017} for this kind of application.
 In \cite{wangAoI} the authors design a radio resource management (RRM) policy to optimize the AoI in vehicular networks that satisfy the demand for real-time and reliable communications. 
 In \cite{kaul2011minimizing} authors minimize the age of status updates sent by vehicles over a carrier-sense multiple access (CSMA) network using the gradient descent scheme.
 The authors of \cite{9562190} investigate the problem of AoI-aware RRM for a platooning system. 
 To solve the corresponding optimization problem, they use the \textcolor{black}{multi-agent reinforcement learning (MARL)} algorithm for distributed resource allocation. 
 
 \textcolor{black}{As can be seen, no study has been conducted to investigate the impact of RRM on path planning validation. However, this is essential to have accurate and real-time information when planning the optimal path for each CV. Accordingly, RRM provides traffic incident messages that comply with the above requirements.}
In our work, we model capacity estimation error utilizing AoI levels, followed by formulating an optimization problem to minimize the AoI of vehicles to validate path planning. 
\\ \\
The contributions of this work lies in the following key items:
\\ $\bullet$ We formulate an optimization problem for each CV to minimize the AoI.
\\ $\bullet$ A model is developed to describe the spectrum access of multiple CVs as a multi-agent problem and utilize the deep deterministic policy gradient (MADDPG) \cite{MADDPG} algorithm.
\\ $\bullet$ We update urban capacities estimation errors according to different AoI values in path planning simulation.
\\ $\bullet$ Numerical experiments demonstrate that the average AoI quantity is maintained within a range of 5-10 milliseconds and fast convergence occurs. Accordingly, path planning performance has significantly improved in terms of travel time (TT) and volume over capacity V/C. \\
\indent \textcolor{black}{The rest of the paper is organized as follows. In Section \ref{II}, the system model is described. We present multi-agent RL-based V2X RRM design in Section \ref{III}. Section \ref{IV} provides the simulation results and concluding remarks follow in Section \ref{V}.}
\vspace{-0.1cm}
\section{SYSTEM MODEL AND PROBLEM FORMULATION }\label{II}
\begin{figure*}
	\centering
	\includegraphics[width=\textwidth, height=8cm]{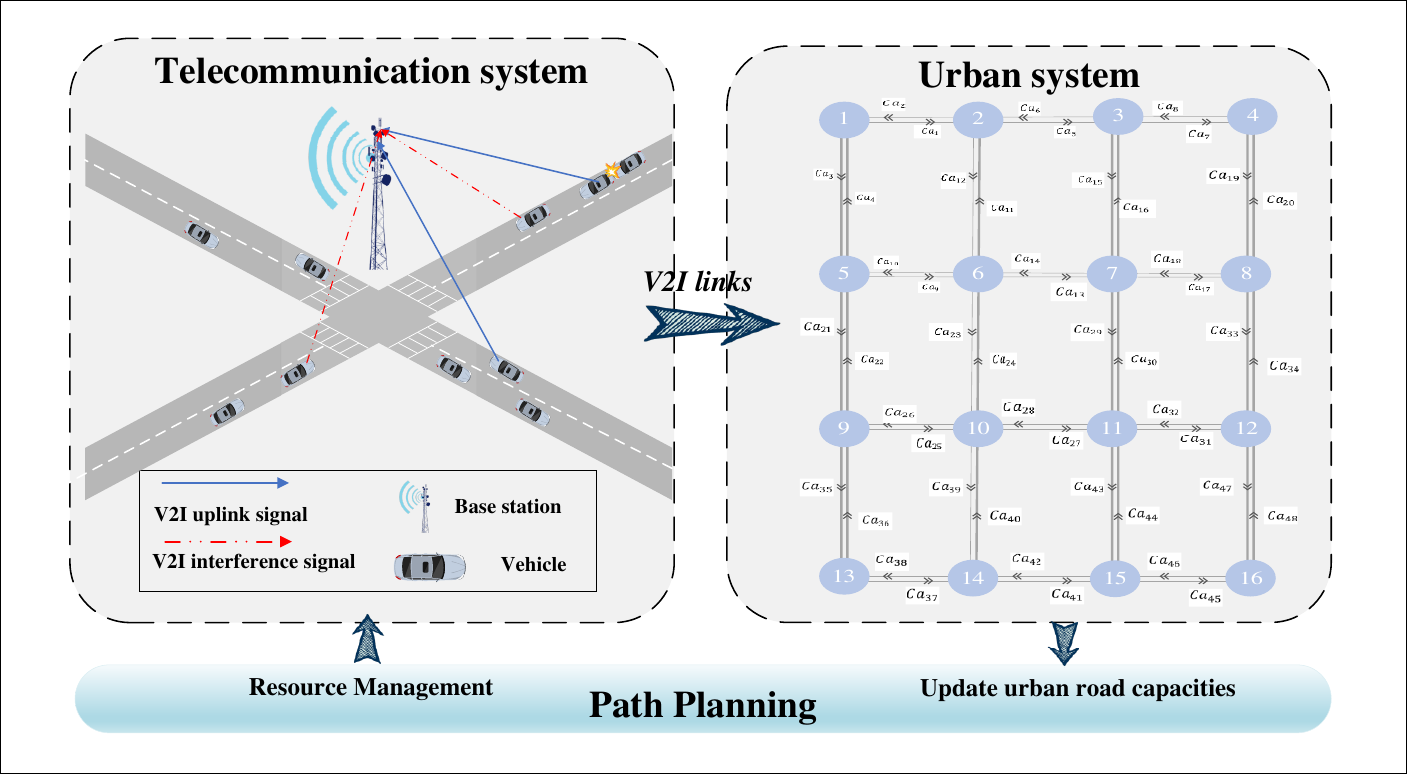}
	\caption{Radio resource management system model  }
	\label{system model} 
\end{figure*}
\subsection{Network model}
As depicted in Fig.~\ref{system model}, our study is based on the utilization of cellular V2X technology for vehicular communication which consists of one BS and multiple CVs. The BS is located in the center of the environment and is equipped with a single antenna, while CVs are also equipped with a single antenna. The BS is mainly responsible for managing the optimal path by collecting data from all network nodes and building a global traffic information graph of all CVs on the roads. For this purpose, the BS updates the road's capacity with the received V2I information from CVs. 
\vspace{-0.4cm}
\subsection{ Radio system model formulation}
We assume that $\mathcal{V}=\{v_1, v_2, \ldots, v_V\}$, with $|\mathcal{V}|=V$ indicates the set of CVs, $V$ represents the total number of CVs. $\mathcal{L}=\{l_1,l_2,\dots,l_L\}$, with $|\mathcal{L}|=L$ indicates the set of roads, where $L$ represents the total number of roads in urban graph model. The time is discretized into equal scheduling slots of length $\Delta t$ with index $t$. The system bandwidth is divided into orthogonal sub-channels of size $W_n$, where indexed by $n \in \mathcal{N}$, and $\mathcal{N}$ is the set of sub-channels. According to the proposed scheme, CV $v$ in road $l$ exchanges highway traffic data 
with the BS in the center of crossroad via V2I links. Orthogonal frequency division multiplexing (OFDM) is used to handle frequency selective wireless channels. We assume that the channel fading is independent across different sub-channels and remains constant within one sub-channel. We model the V2I channel gain of CV to BS on sub-channel $n$ during one $\Delta t$ $t$ as,
\vspace{-0.1cm}
\begin{align}
	h_{v}^{t}[n]=\alpha_{v}^{t} g_{v}^{t}[n],
\end{align}
It is worth noting that $\alpha_{v}^{t}$ and $g_{v}^{t}[n]$ represent the large-scale fading effect, which is the result of path loss and shadowing, and the small-scale fading, respectively. In addition, we define a binary variable $\rho_{v}^{t}[n] \in$ $\{0,1\}$ that indicates whether sub-channel $n$ is allocated to CV $v$ at time slot $t$, $\rho_v^t[n]=1$, otherwise $\rho_v^t[n]=0$. We can describe the instantaneous rates 
between CV $v$ and the BS according to the Shannon capacity formula as follows:
\vspace{-0.1cm}
\begin{align}
	\nonumber
	&\mathcal{C}_v^t[n]= W_n\log_{2}\left(1+\frac{\rho_{v}^{t}[n] p_{v}^{t}[n] h_{v}^{t}[n]}{I_{v}^{t}[n]+\sigma^{2}_n}\right), \\
	&I_{v}^{t}[n]=\sum_{v^{\prime} \in V} \rho_{v^{\prime}}^t[n] p_{v^{\prime}}^t[n] h_{v^{\prime}}^t[n], \: v^{\prime} \neq v, 
\end{align}
where $p_{v}^{t}[n]$ is the power level used by CV $v$ on sub-channel $n$ for transmitting, 
$h_{v^{\prime}}^{t}[n]$ is the interfering channel to BS from CV $v^{\prime} \in \mathcal{V}$, and $I_{v}^{t}[n]$ corresponds to the total interference power. As previously described, the CV has to ensure timely communication with the BS to exchange any dynamic incidents. As a result, we note $A_{v}^{t}$ as the AoI of CV $v$ up to the beginning of the scheduling slot t, it refers to the period of time that has passed since the last successful V2I communication. It is worth noting that this communication inform BS of dynamic traffic capacities. The AoI of CV $v$ evolves according to the following recurrence relation:
\vspace{-0.1cm}
\begin{align}
	A_{v}^{t+1}= \begin{cases}\Delta t, & \text { if } \rho_{v}^{t}[n] \cdot \mathcal{C}_{v}^{t}[n] \geq \mathcal{C}_{v}^{\min } \\ A_{v}^{t}+\Delta t, & \text { otherwise }\end{cases},
\end{align} 
In this case, $\mathcal{C}_{v}^{\min }$ is the minimum capacity required for V2I communication.
\vspace{-0.4cm}
\subsection{Path planning}
\textcolor{black}{For assigning the traffic demand between different origins and destinations pairs of the network to available routes, one can take either user equilibrium (UE) or system optimization (SO) approach. In UE condition, no driver can unitedly reduce his/her travel time and according to Wardrop’s principles, all the available routes between specific origin-destination pairs have the same travel time. This condition naturally exists in the network because all drives tend to optimize their travel time by changing routes. Under this condition the total travel time of the network may not be minimum. On the other hand, the objective function of SO problem is the total travel time spent in the network and the traffic flow patterns that solves this problem, minimizes the objective function. In our case study, the number of assigned CVs to each route with both approaches are the same due to consistency of travel times in each system iteration. For this reason, we take the UE assignment approach to simulate the network and use the Dijkstra algorithm for finding the shortest path based on travel time according to the bereau of public roads (BPR) equation as below:}
\vspace{-0.2cm}
\begin{align}
	T_l(F_l) = T_l(0)  \left[1 + \alpha_u \left(\frac{F_l}{Ca_l}\right)^{\beta_u} \right],
\end{align}
where $l$ is an index of road, $T_l(0)$, $T_l$, $F_l$ and $Ca_l$ are free flow time, travel time, traffic flow, and capacity of the road, respectively. $\alpha_u$ and $\beta_u$ are two parameters of this equation that are mostly regarded as $0.15$ and $4$\textcolor{black}{\cite{Skabardonis}}. At last, the Frank-Wolfe (FW) algorithm is applied to solve an iterative first-order optimization. As described earlier, different incidents can affect capacity of a road. In this regard, we denote $Ca_l$ as the real capacity of road,
\vspace{-0.2cm}
\begin{align}
	\nonumber
	&Ca_l= \hat{Ca}_l+\Delta ca,\\
	& \Delta ca=\left(-(A_v)/A_{v,\mathfrak(R)}^{\text{max}} \right)\Delta m ,
\end{align}
where $\hat{Ca}_l$ and $\Delta ca$ denote the estimated capacity of each road, and deviation from the actual value, respectively. The BS which is responsible for path planning can estimate the roads capacities more accurately with fresh data collected by V2I communications. $A_v$ is a measure of the freshness of information. $A_{v,\mathfrak(R)}^{\text{max}}$ is the maximum AoI between CV $v$ and the BS, and $\Delta m$ is the maximum amount of error in road's capacities estimation. Thus, the optimization problem for each CV $v$ can be expressed as follows:
\vspace{-0.2cm}
\begin{align}		\label{objective function} 
		\min _{\boldsymbol{\rho}, \boldsymbol{p}}&\left\{\frac{1}{T}\sum_{t=1}^{T} A_{v}^{t}\right\}\\
		\text { s.t. } \: &C 1: \mathcal{C}_{v}^{t}[n] \geq \mathcal{C}_{v}^{\min }, \quad \forall v \in \mathcal{V},\forall n \in \mathcal{N},\\
		\: &C 2: \rho_{v}^{t}[n] \in\{0,1\}, \quad \forall v \in \mathcal{V}, \forall n \in \mathcal{N}, \\ 
		\: &C3: \sum_{n \in \mathcal{N}} \rho_{v}^{t}[n] \leq 1, \quad \forall t \in \mathcal{N}, \forall v \in \mathcal{V}, \\
		\: &C4: p_{v}^{t}[n] \leq p_{v}^{\max }, \quad  \forall v \in \mathcal{V},\forall n \in \mathcal{N}.
\end{align}

The objective is to minimize the 
\textcolor{black}{average} AoI for every CV within every $T$ seconds, which leads to the correct estimation of the capacity and its optimal use. Constraint C3 indicates that each CV is only permitted to access one sub-channel per time slot. C4 mandates that CV $v$'s transmit power remain below its maximum value $p_{v}^{\max}$. In the optimization problem, the sub-channel selection parameter is a binary variable. The objective function is also non-convex. Therefore, the optimization problem is an NP-hard combinatorial optimization problem that is difficult to solve. To address the complexities of the proposed optimization problem, we investigate the state-of-the-art MADDPG method.
\section{MADDPG based solution}\label{III}

In a multi-agent environment, agents are intended to maximize their policy functionalities, as outlined below: 
\begin{align}
	\label{rewardd}
	\max _{\pi} {J}^{v}\left(\pi^{v}\right), \: v \in \mathcal{V}, \: \pi^{v} \in \Pi^{v},
\end{align}
where ${J}^{v}\left(\pi^{v}\right)=\mathbb{E}\left[\sum_{t=0}^{\infty} \gamma^{t} \tilde{r}_{t+1}^{v} \mid s_{0}^{v}\right]$ is our conditional expectation, $\pi^{v}$ describes the policy of CV $v$, additionally, the set $\Pi^{v}$ includes all feasible policies that are available for CV $v$. In the proposed model, a CV interacts with the vehicular network environment takes actions in accordance with its policies in order to address the optimization problem, \ref{objective function}. The agents strive to improve their reward \ref{rewardd}, which is linked to the objective function. At any time $t$, the CV observes a state, $s^t$, and takes appropriate action, $a^t$. As a result of the CV's selected action, the environment transitions into a new state $s^{t+1}$ and a reward is given to the CV. As a result of our proposed system model, we define the state space, $\mathcal{S}$, the action space,  $\mathcal{A}$, and the reward function, $r^t$, as follows:\\
\indent \textbf{State space:} Multiple components comprise the state observed by the CV at time slot $t$: the instant channel information between CV $v$ and the BS, $h_{v}^{t}[n]$ for all $n \in \mathcal{N}$, \textcolor{black}{the previous interference from other CVs to CV $v$, and the AoI of CV $v$, $A_v^t$.}
Accordingly, the state space of CV $v$ is 
\begin{align}
	\mathbf{s}_v^t=[h_{v}^{t}[n], I_v^{t-1}[n],A_v^t ],\: v\in\mathcal{V}.
\end{align}
\indent \textbf{Action space:} The action of each CV $v$ is defined as $\mathbf{a}_{v}^{t}=\left\{\rho_{v}^{t}, p_{v}^{t}\right\}$. Earlier, we mentioned that $\rho_{v}^{t}[n]$ being 0 or 1 indicates if sub-channel $n$ has been selected for a give $v$ and $t$, and $p_{v}^{t}$ represents the power at time $t$ for CV $v$ on sub-carrier $n$. Using the deep deterministic policy gradient (DDPG) method, the agent is able to select any power within the range of 0 to $p_{v}^{\max }$. This is the advantage of policy gradient method where we can apply continuous actions spaces and can converge to more accurate results than conventional \textcolor{black}{deep Q learning (DQNs)} in which the power has to be discretized.\\
\indent \textbf{Reward space:} In this study, we use a learning problem that involves agents receiving two reward, a global reward for their cooperation and an individual reward for their performance. Two critics will be trained in the following functions as part of this framework: global centralized critics use observations and actions of all agents as inputs to estimate the global team reward. The local critics, which are specific to each agent, estimate the rewards based on the agent's local observations and actions. The goal is to optimize both global and local rewards simultaneously through the policy. We first discuss \textcolor{black}{MADDPG}
framework and then design our rewards. 
\\ \indent We use $\pi$ to indicate the set for all CVs policies where CV $v$ policy is denoted by $\pi^{v}\left(\pi^{v} \in \boldsymbol{\pi}=\left\{\pi^{1}, \ldots, \pi^{V}\right\}\right)$ with parameters $\tilde{\theta}^{v}$, $\tilde{Q}^{v}$ for CV critic (Q-function) with parameter $\tilde{\phi}^{v}$ and $\tilde{Q}_{G}$ for global Q-function with parameter $\tilde{\psi}$. In the first step, we form a neural network, and in the second step, we discuss gradient policies. The number of layers in neural networks for each agent action, critic, and global critic are $L_{\pi}^{v}, L_{q}^{v}$, and $L_{G}$, respectively. As a result of the above information, we can develop our neural network in this manner $\tilde{\Theta}_{i}=\left(W_{\pi}^{(1)}, \ldots, W_{\pi}^{(L)}\right)$, $\tilde{\Phi}_{q}=\left(W_{q}^{(1)}, \ldots, W_{q}^{(L)}\right)$, and $\tilde{\Psi}_{G}=\left(W_{G}^{(1)}, \ldots, W_{G}^{(L)}\right)$ as actor, CV critic and global critic. It can be concluded from the above that the gradient policy has the following characteristics:
\begin{align}
	\nabla_{\theta^{v}} {J}^{v}=\mathbb{E}\left[\left.\nabla_{{\tilde{\theta}^v}} \pi^{v}\left(a^{v} \mid s^{v}\right) \nabla^{a^v} Q_{\pi}^{v}(\mathbf{s}, \mathbf{a})\right|_{{a^{v}}={\pi^{v}}(s^{v})}\right],
\end{align}
where $s=(s_1, \dots, s_N)$ and $a=(a_1, \dots, a_N)$ are the total state and action vectors. In order to approximate Qfunction for CV $v$, all actions and states in $Q_{\pi}^{v}(\mathbf{s}, \mathbf{a})$ are integrated as inputs. To develop the policy gradient, we utilize two critic networks, which are reformulated as follows:
\begin{align}\label{policy gradient}
	\nabla_{\theta^{v}} {J}^{v} &=\underbrace{\mathbb{E}_{\mathbf{s}, \mathbf{a} \sim \mathcal{D}}\left[\nabla_{\theta^{v}} \pi^{v}\left(a^{v} \mid s^{v}\right) \nabla_{a^{v}} Q_{G}^{\psi}(\mathbf{s}, \mathbf{a})\right]}_{\text {Global Critic }}\\ \nonumber
	&+ \underbrace{\mathbb{E}_{s^{v}, a^{v} \sim \mathcal{D}}\left[\nabla_{\theta^{v}} \pi^{v}\left(a^{v} \mid s^{v}\right) \nabla_{a^{v}} Q_{\phi^{v}}^{v}\left(s^{v}, a^{v}\right)\right]}_{\text {CV critic }},
\end{align}
where $a^v=\pi^v(s^v)$ is the action that agent $v$ chooses under policy $\pi^v$ and $\mathcal{D}$ is the replay buffer. In (\ref{policy gradient}), two terms are used: the first shows the global critic, which receives the actions and states of all CVs and estimates the global Q-function using the global rewards $\tilde{r}_{G}$, and the second shows the critic of each CV, which receives only its own actions and states. The loss function of the global critic is updated by using
\vspace{-0.2cm}
\begin{align}
	\mathcal{L}(\psi)=\mathbb{E}_{\mathbf{s}, \mathbf{a}, \mathbf{r}, \mathbf{s}}\left[\left(Q_{G}^{\psi}(\mathbf{s}, \mathbf{a})-y_{G}\right)^{2}\right],
\end{align}
where $y_{G}$ represents the target value of the estimation as follows: 
\begin{align}
	y_{G}=\tilde{r}_{G}+\left.\gamma Q_{G}^{\psi}\left(\mathbf{s}^{\prime}, \mathbf{a}^{\prime}\right)\right|_{a^{\prime v}=\pi^{\prime v}(s^{\prime v})}.
\end{align}
\\ The target policy is $\pi^{\prime}=\left\{\pi^{\prime 1}, \ldots, \pi^{\prime V}\right\}$. It is parameterized with $\boldsymbol{\theta}^{\prime}=\left\{\theta^{\prime 1}, \ldots, \theta^{\prime V}\right\}$, and the CV loss function and its target update are as follows:
\begin{align}
	\mathcal{L}^{v}\left(\phi^{v}\right)=\mathbb{E}_{\mathbf{s^{v}}, \mathbf{a^{v}}, \mathbf{r^{v}}, \mathbf{s^{\prime v}}}\left[\left(Q_{\phi^{v}}^{v}\left(s^{v}, a^{v}\right)-y^{v}\right)^{2}\right],
	\label{loss}
\end{align} 
\begin{align}
	y^{v}=r^{v}+\left.\gamma Q_{\phi^{\prime v}}^{v}\left(s^{\prime v}, a^{\prime v}\right)\right|_{a^{\prime v}=\pi^{\prime v}(s^{\prime v})}.
	\label{target}
\end{align} 
 On the basis of the results presented in \cite{hybrid}, we replace the global critic with the twin delayed (TD3) deterministic policy gradient in order to overcome the problem of overestimation and suboptimal policies resulting from errors in function approximation. In this regard, the gradient of policy is as follows:\\
\vspace{-0.3cm}
\begin{align} 	\label{td3}
	\nabla_{\theta^{v}} {J}^{v}&=\underbrace{\mathbb{E}_{\mathbf{s}, \mathbf{a} \sim \mathcal{D}}\left[\nabla_{\theta^{v}} \pi^{v}\left(a^{v} \mid s^{v}\right) \nabla_{a^{v}} Q_{G_i}^{\psi_i}(\mathbf{s}, \mathbf{a})\right]}_{\text {TD3 Global Critic }}\\ \nonumber
	&+ \underbrace{\mathbb{E}_{s^v, a^v \sim \mathcal{D}}\left[\nabla_{\theta^v} \pi^{v}\left(a^{v} \mid s^{v}\right) \nabla_{a^v} Q_{\phi^v}^{v}\left(s^{f}, a^{f}\right)\right]}_{\text {CV critic }},
\end{align} 
In (\ref{td3}), the twin global critics are updated as follows:
\begin{align}
	\mathcal{L}\left(\psi_{i}\right)=\mathbb{E}_{\mathbf{s}. \mathbf{a}, \mathbf{r}, \mathbf{s^\prime}}\left[\left(Q_{G_i}^{\psi_i}(\mathbf{s}, \mathbf{a})-y_{G}\right)^{2}\right],
\end{align}
where $y_G$ is updated as follows:
\begin{align}
	y^{v}=r^{v}+\left.\gamma(1-\tilde{d}) \min _{i=1,2} Q_{\phi_i^{\prime
			v_i}}^{v_i}\left(s^{\prime v}, a^{\prime v}\right)\right|_{a^{\prime i}=\pi^{\prime i}(s^{\prime i})}.
\end{align}
The agents' critic networks are updated by \ref{loss} and \ref{target} and there is a detailed description of the MADDPG algorithm in \ref{algorithm}. In TD3, the main idea is to wait for the convergence of value estimates before updating the policy.  We are now able to design our rewards, each agent should keep updated connections with the BS and maintain a minimal level of AoI. The global reward should assist the CVs in choosing the proper sub-channel and power to minimize interference between the CVs. To this end, we design individual agent and global reward as 
\begin{align}
	& r_{\ell}^{v}=-{\frac {A_{v}^{t}}{V}+0.05G\left(\mathcal{C}_{v}^{t}-\mathcal{C}_{v}^{\text {min }}\right)},\\
    & r_{g}=\frac{1}{V} \sum_{v \in \mathcal{V}} r_{\ell}^{v},
\end{align}
where $G(x)$ is the stepwise function used to satisfy  $C3$.

\begin{algorithm}
	\scriptsize
	\renewcommand{\arraystretch}{0.4}
	\caption{Two time frame modified MADDPG}
	\label{algorithm}
	Generate CVs and establish an environment
	\\ \textbf{Input:} For each agent, enter the number of states and actions
	\\ Initiate the main global critic networks  $Q_{\phi_1}^v$ and  $Q_{\phi_2}^v$
	\\ Initiate the target global critic networks of $Q_{\phi_1}^{\prime v}$ and  $Q_{\phi_2}^{\prime v}$
	\\ Set up the policy and critic networks for each agent
	\\\For{each episode}{
		Provide an update on CV locations and respective channel gains
		\\ Reset the payload of the V2I message and maximum transmission time $T$ to ${100 \mathrm {~ms}} $
		\\\For{each timestep $t$}{
			\For{each agent $v$}{
				observe $s_v^t$ and Choose an action $a_v^t$
			}
			${s_t=[s_t^1,\cdots, s_t^V]}$ , ${a_t=[a_t^1,\cdots, a_t^V]}$
			\\ Receive global and local rewards, $r_g^t$ and $r_l^t$
			\\ Store $(s^t, a^t, r_l^t, r_g^t, s^{t+1})$ in replay buffer $\mathcal{D}$
		}
		Sample minibatch of size $S$, $(s^j, a^j, r_l^j, r_g^j, s^{\prime j})$, from the replay buffer $\mathcal{D}$ 
		\\  $y_g^j = r_g^j + \gamma \min_i Q_{\psi_i^\prime}^{g_i}(s^{\prime j}, a^{\prime j})$
		\\ Provide an update to global critics by minimizing the loss:
		\\ $\mathcal{L}(\psi_i) = \frac{1}{S} \sum_j \left\{{\left( Q_{\psi_i}^{g_i}(s^j, a^j)-y_g^j\right)}^2\right\}$ 
		\\ Update target parameters: $\psi_i^\prime \gets \tau \psi_i + (1-\tau)\psi^\prime_i$
		\\\If{episode mode d}{
			Train local critics and actors 
			\\\For{each agent $i$}{
				Provide an update to local critics by minimizing the loss:
				\\ $\mathcal{L}(\phi_i) = \frac{1}{S} \sum_j \left\{{\left( Q_{\phi_i}^{i}(s^j_i, a^j_i)-y_i^j\right)}^2\right\}$
				\\ Provide an update to local critics: 
				\\ $\nabla {J} \theta_i \approx \frac{1}{S} \sum_j \nabla \theta_i \pi_i \left(a_i \mid s_i^j\right) \nabla_{a_i} Q_{\psi_1}^{g_1} (s^j, a^j)  +  \nabla _{\theta_i} \pi_i (a_i \mid s_j^j) \nabla_{a_i} Q_{\phi_i}^i(s_i^j, a_i^j)$
				\\ Provide an update to target networks parameters 
				\\ $\theta_i^{\prime} \gets \tau \theta_i + (1-\tau)\theta_i^{\prime}$ , $\phi_i^\prime \gets \tau \phi_i + (1-\tau)\phi_i^\prime$		
			}
		}
		
	}
	\vspace{-0.2 cm }
\end{algorithm}

\section{SIMULATION RESULTS}\label{IV}

\begin{table}
	\centering
	\caption{Parameters for simulation}
	\label{table:tableofnotations}
	\scalebox{0.9}{
		\begin{tabular}{ll}
			\vspace{-0.2 cm}
			& \vspace{-0.3 cm} \\
			\hline
			Parameter & Value \\
			\hline
			Frequency of carrier & $2 \mathrm{GHz}$ \\
			Number of Resource blocks & 3 \\
			Each Resource block's bandwidth & $180 \mathrm{kHz}$ \\
			Number of CVs in RRM simulation & 20 \\
			CVs Speed & $10-14 \mathrm{~m} / \mathrm{s}$ \\
			BS and CVs antenna heights & $25,1.5 \mathrm{~m}$ \\
			BS and CVs antenna gains & $8,3 \mathrm{dBi}$ \\
			BS and CVs receiver noise figure & $5,9 \mathrm{~dB}$ \\
			CVs mobility model & Urban case of A.1.2 \cite{36.8851} \\
			CVs maximum power & $30 \mathrm{dBm}$ \\
			Noise power $\sigma^2$& $-114 \mathrm{dBm}$ \\
			Time constraint of V2I message dissemination, $T$, & $100 \mathrm{~ms}$ \\
			V2I links path loss model & $128.1+37.6 \log_d(10)$ \\   
			Shadowing distribution & Log-normal \\
			Standard deviation of shadowing for V2I links & $8 \mathrm{~dB}$ \\
			Decorrelation distance for V2I links & $50 \mathrm{~m}$ \\
			V2I path loss/shadowing update & Every $100 \mathrm{~ms}$\cite{36.8851} \\
			V2I fast fading update & Every $1 \mathrm{~ms}$\cite{36.8851} \\
			Fast fading & Rayleigh fading \\ 
			\hline
			Replay buffer size for experience & 100000 \\
			Size of the mini batch & 64 \\
			Hidden layer number and size of actor networks & $2 / 1024,512$ \\
			Hidden layer number and size of critic networks  & $3 / 1024,512,256$ \\
			Learning rate of critic/actor networks & $0.001 / 0.0001$ \\
			Factor of discount & $0.99$ \\
			Target networks soft update parameter & $0.0005$ \\ \hline
			Number of CVs in urban simulation & $240$ \\
			Normal roads' capacity& $30 \mathrm{~veh/h}$ \\
			Roads' length & $250, 433 \mathrm {~m}$ \\
			Free flow time & $18,31 \mathrm {~s}$\\
			Number of roads & $48$ \\
			$\alpha_u$ & $0.15$ \\
			$\beta_u$ & $4$ \\
			\hline
			\vspace{-0.6 cm}
		\end{tabular}
	}
\end{table}

\begin{figure}[t]
	\includegraphics[width=0.87\linewidth]{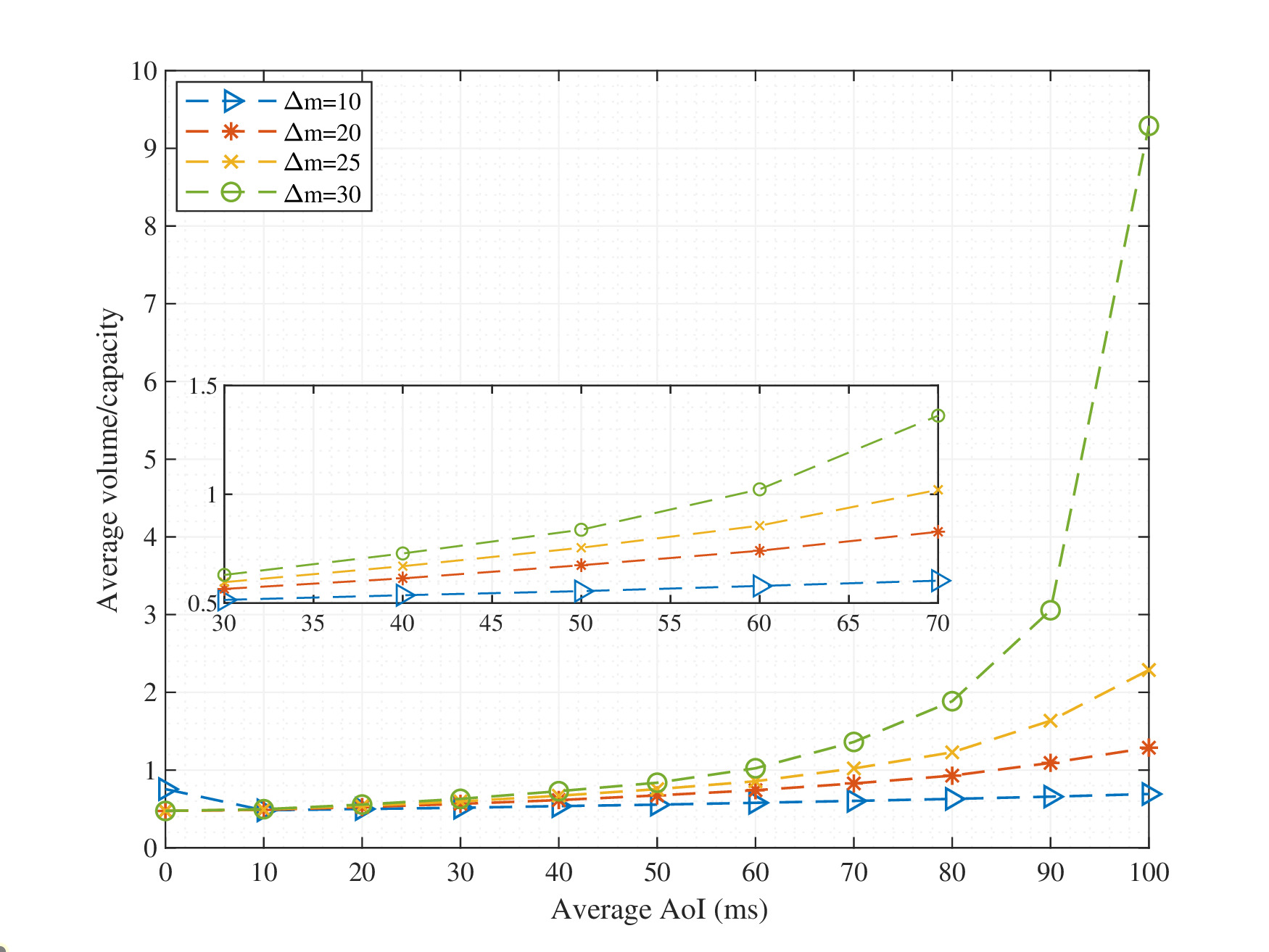}
	\centering
	\caption{Average V/C with different value of $\Delta m$}
	\label{voverc} 
\end{figure}

\begin{figure}[t]
	\includegraphics[width=0.87\linewidth]{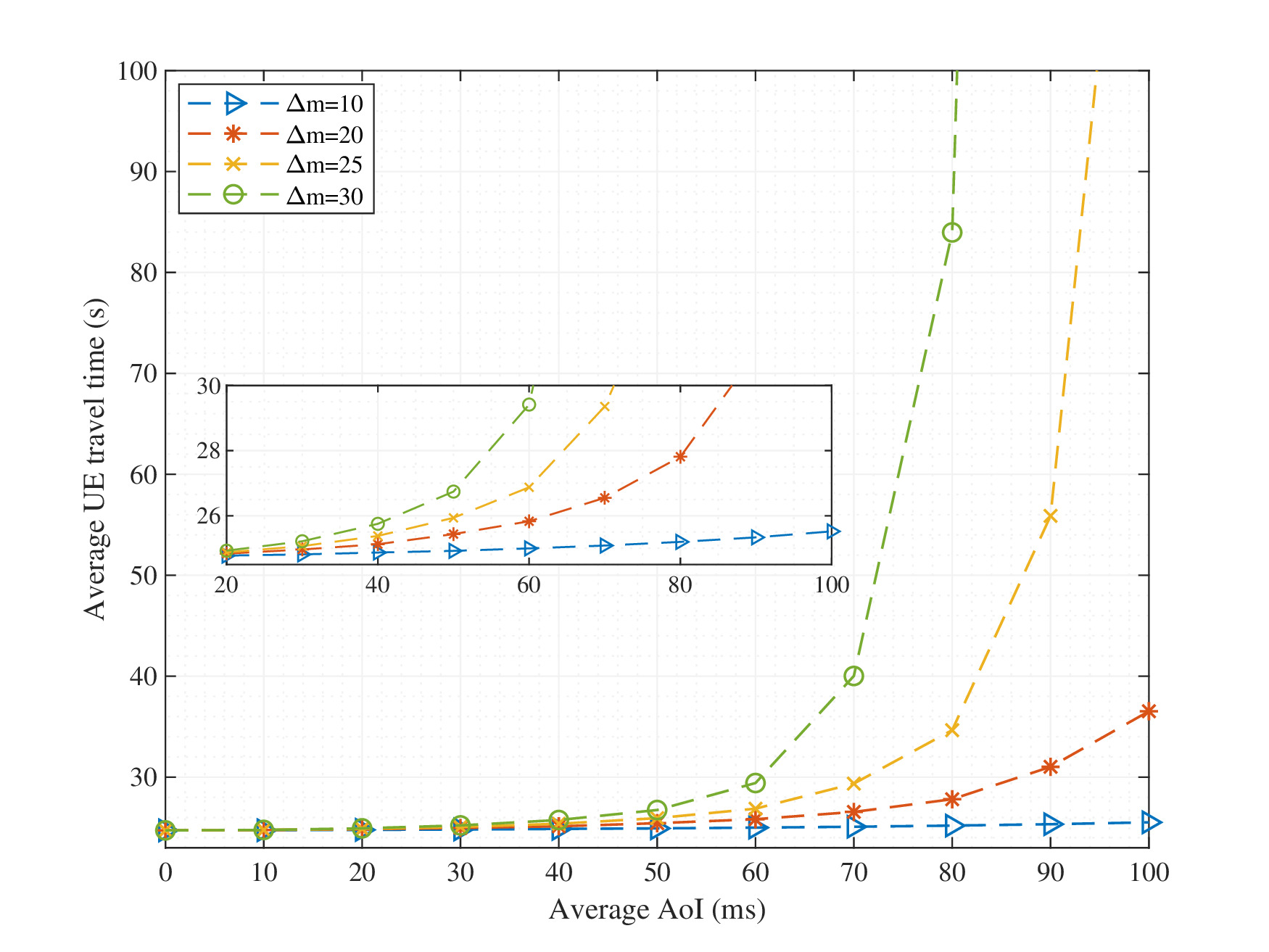}
	\centering
	\caption{Average roads travel times with different value of $\Delta m$}
	\label{tt} 
\end{figure}

To evaluate the effectiveness of our formulation, we analyzed a network of 16 nodes, 48 roads, and 240 origin-destination pairs. In this scenario, CV flow of 240 CVs per hour is considered, and the capacity of each road is assumed to be 30 CVs per hour. A summary of the simulation parameters can be found in \ref{table:tableofnotations}. The simulation results for CVs driving in this scenario are depicted in Fig.~\ref{tt} and~\ref{voverc}. We plotted the roads' average travel time and a variable of the roads' quality of service named volume over capacity (V/C) ratio against average AoI for four different values of $\Delta m$. V/C is a measure of capacity sufficiency, that is, whether or not the physical geometry and path planning system provides sufficient capacity for the network flow. In the case of low traffic flow in the network, the V/C ratio tends to zero which shows an ideal state of the network. In contrast, as the traffic flow in the network increases, the V/C ratio tends to 1. In the case of an overcrowded state in the network, this ratio passes 1 and the roads' quality of service is very poor.

Fig.~\ref{tt} depicts the average roads travel times versus average AoI values. We can figure out that the average travel time of a CV grows as the average AoI increases. An increase in AoI causes more error in road capacities estimation and, as a result, we face invalid path planning. Moreover, it is shown that in lower values of $\Delta m$, the growth of average travel time is more negligible compared to high values of $\Delta m$.

Figure~\ref{voverc} plots the values of the V/C ratio against average AoI values. These plots start from a value in the range of zero and 1. As the average AoI increases, the path planning framework fails to update the road capacities accurately and the volume of roads grows, and congestion forms. As a result, the volume of roads gradually grows to the capacity and the V/C ratio passes $1$. As mentioned earlier, in lower values of $\Delta m$, the increase in the V/C ratio is smaller compared to higher values of $\Delta m$. 
\begin{figure}[t]
	\includegraphics[width=0.87\linewidth]{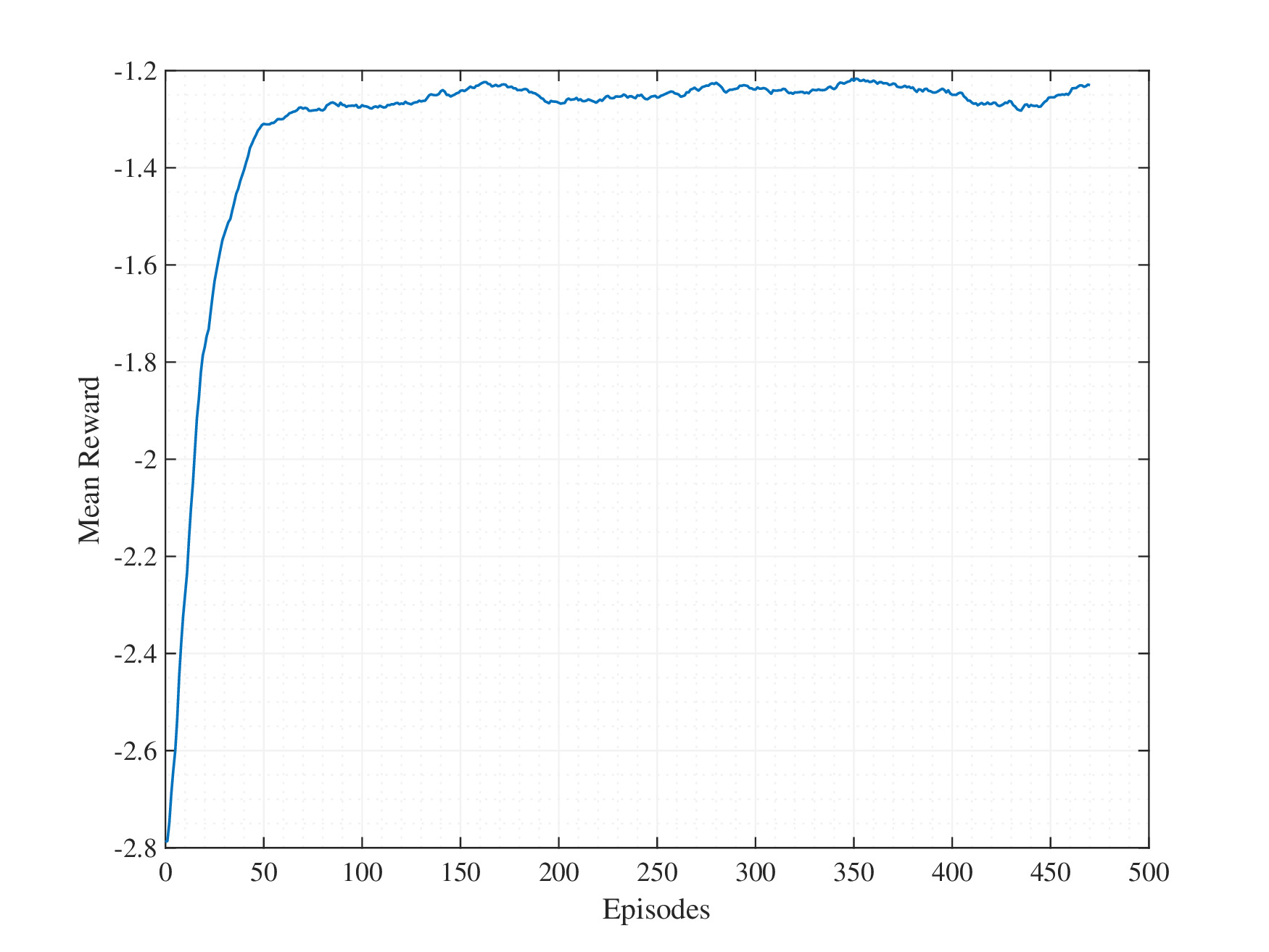}
	\centering
	\caption{Average reward over learning episodes.}
	\label{reward} 
\end{figure}
\begin{figure}[t]
	\includegraphics[width=0.87\linewidth]{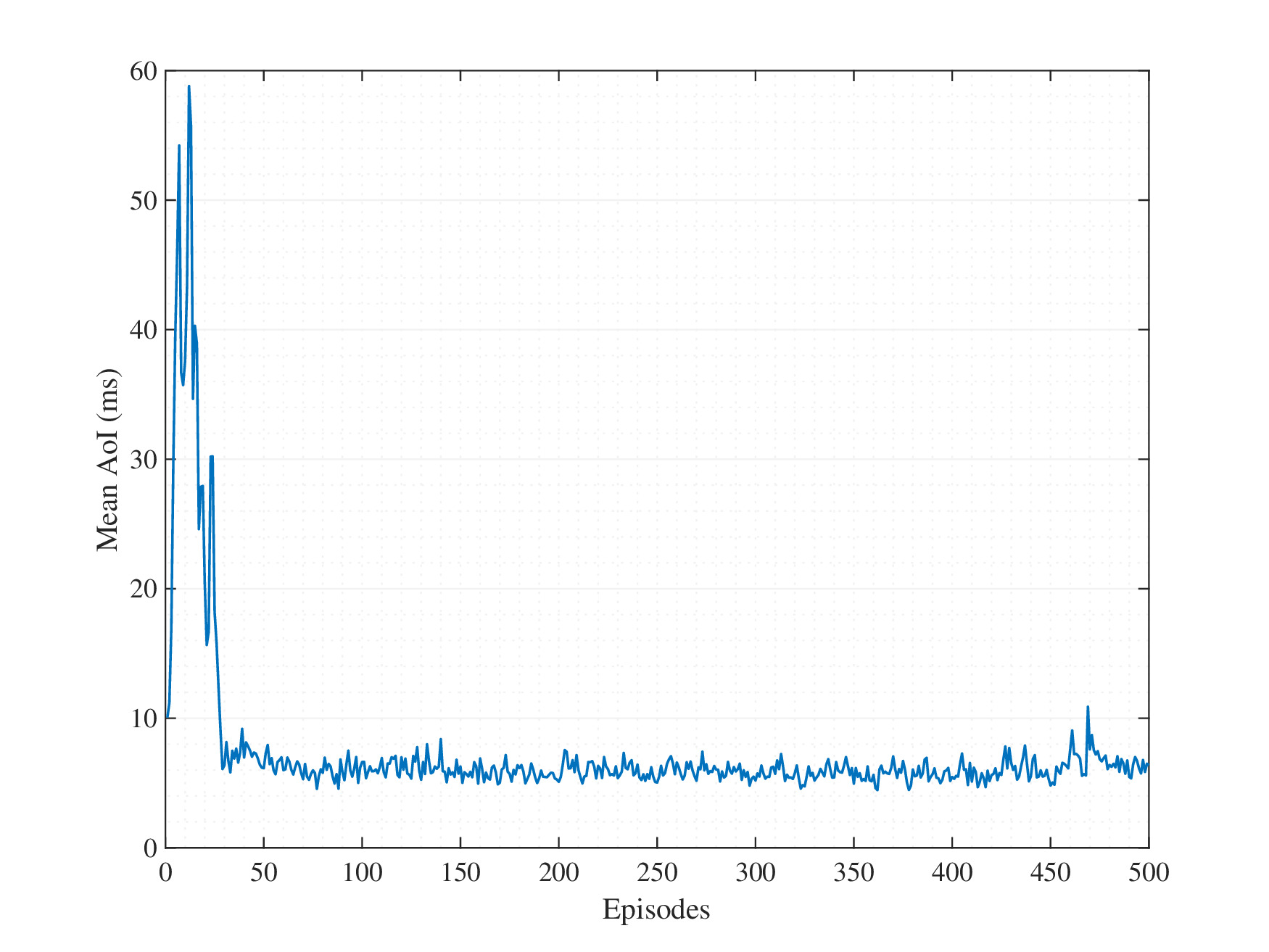}
	\centering
	\caption{Average AoI over learning episodes.}
	\label{AverageAoI} 
\end{figure}
Fig.~\ref{reward} shows the convergence of proposed MADDPG framework reward. In the proposed framework, individual and global rewards are maximized simultaneously for all agents, resulting in a stable performance. From Fig.~\ref{AverageAoI}, it is evident that as the number of episodes increases, the AoI quantity decreases. The utilized \textcolor{black}{MADDPG} 
framework maintains the average AoI values from 60 to below 10$\mathrm {~ms}$.
\vspace{-0.2 cm}

\section{CONCLUSION}\label{V}
As presented in this paper, a joint path planning scheme and radio resource allocation was developed for vehicular systems in
order to minimize AoI of CVs while updating urban road capacities for effective path planning. In our work, we were able to confirm that the presented RRM scheme is highly robust and effective in terms of enticing CVs
to improve the model of their systems by minimizing their AoI, thereby reducing the error in estimating the capacity. The future work will explore more complicated system models including other technologies such as UAVs to monitor the urban environment for dynamic incidents.

\vspace{-0.18 cm}
\bibliographystyle{ieeetr}
\bibliography{Citation}
\end{document}